\documentclass{aa}             % <-- "class" anstelle von "style"
\usepackage{amssymb,graphicx}  % <-- "graphicx" fuer Bilderhat "amssymb"
\newcommand{\promille}{%
     \relax\ifmmode\promillezeichen
           \else\leavevmode\(\mathsurround=0pt\promillezeichen\)\fi}
   \newcommand{\promillezeichen}{%
     \kern-.05em%
     \raise.5ex\hbox{\the\scriptfont0 0}%
     \kern-.15em/\kern-.15em%
     \lower.25ex\hbox{\the\scriptfont0 00}}
\begin{document}
\thesaurus{ 02.01.2; 08.14.2; 08.09.2; 13.25.5; 08.03.5}
\title{WZ Sagittae - an old dwarf nova}
\author{
E.~Meyer-Hofmeister \inst{1},
F.~Meyer \inst{1},
B. F. Liu \inst{2}
}
\offprints{Emmi Meyer-Hofmeister}
\institute{
Max-Planck-Institut f\"ur Astrophysik, Karl
Schwarzschildstr.~1, D-85740 Garching, Germany
\and
Yunnan Observatory, Academia Sinica. P.O.Box 110, Kunming 650011, China
}
\date{Received date; accepted date}
\maketitle
\markboth{WZ Sagittae}{E.\ Meyer-Hofmeister et al.}
\begin{abstract}

We model the evolution of the accretion disk of WZ
Sagittae during the long quiescence. We find that the large amount of mass
in the disk derived from the outburst luminosity is a severe
constraint and demands values of $\alpha_c$ $\approx$ 0.001 in
contradiction to some recent suggestions. 
We include in our computations the formation of an inner disk
hole and the growth of the disk due to redistribution of angular
momentum. We find a new mode of disk evolution. The disk is
quasi-stationary. Only about half of the mass transfered from the
companion star flows through the disk, the other half is needed to
build up
the steadily growing outer disk. When 
the 3:1 resonance radius is reached the disk growth ends. From then on
all transferred matter flows inward, the surface density increases,
leading to an outburst within a few years. We predict superhumps at low
luminosity during this last phase. We discuss X-rays expected, the
white dwarf mass and distance to WZ Sagittae.

\keywords{accretion disks -- cataclysmic variables --
stars: individual: WZ Sge --  X-rays: stars -- stars: coronae}
\end{abstract}

%-----------------------Introduction---------------------------
\section{Introduction}
WZ Sagittae is a dwarf nova with extremely long time intervals between
outbursts. During the
vvlast outburst superhumps were discovered, which led to the classification
of WZ Sge as a SU UMa type dwarf nova (Patterson et al. 1981).
A review of investigations of WZ Sge in
the past as well as a redetermination of the system parameters is
given by Smak (1993). The
orbital period of 81 minutes is close to the period minimum of 
cataclysmic variables and by standard interpretation puts WZ Sge into a 
late state of evolution of dwarf nova systems. For these close
binaries the distance of secondary star and white dwarf primary
decreases continously due to loss of angular momentum from the system
by gravitational radiation until a period minimum is reached
and it grows again for later times.
Osaki (1996, review) showed that, in the framework of the
thermal-tidal instability model, the decreasing number of
normal outbursts of SU UMa systems between superoutbursts can be
explained as an evolutionary sequence with decreasing mass overflow
rate from the secondary star. WZ Sge is an extreme case in only showing
superoutbursts. Due to its outburst behaviour it was 
estimated that the viscosity parameter in the quiescent state should
be as low as $10^{-3}$ or even $10^{-4}$ (Smak 1993).
Recently two different
suggestions were made, by Lasota et al. (1995),
Hameury et al. (1997) and by Warner et al. (1996)
how the behaviour of WZ Sge could be modeled using a viscosity parameter
in the cool state $\alpha_c$ =0.01,
which is only slightly smaller than what is usually chosen to describe
ordinary dwarf nova outburst cycles. We show with our
computations that the viscosity has to be low, if we take the large
amount of mass in the disk, 1-2\, $10^{24}$g, as a constraint. This mass
was deduced in two ways, from the UV outburst luminosity and from the hot
spot brightness during quiescence (Smak 1993). Also the optical outburst
lightcurve shows the high mass flow rate in the
beginning. SIn addition Smak's value for the distance to WZ Sge is in the lower
range of values given in the literature. Thus a high amount of mass in the
disk seems unavoidable.
 
The evaporation of the inner
disk is an important feature for the evolution of the disk,
because it prevents the early onset of the instability. We include the
physics of the hole formation and find that the hole can either persist
during the long quiescence for systems containing a white dwarf of
large mass or the hole can be closed for small white dwarf mass. 
But evaporation still continues at the inner disk edge close to the white
dwarf. Important for the evolution is also to follow in
detail how the disk grows due to the redistribution of angular momentum
until, for some cases, the 3:1 resonance radius is reached and finally
the onset of the disk instability occurs. We found that during the long
quiescence
the disk can reach a quasi-stationary state. X-rays originate from an
inner disk corona and the thermal boundary layer of the white dwarf.

We give a short description of observational data and
model parameters in Sect. 2. In Sect. 3 we describe the previous 
modelling of the long-term behaviour of WZ Sge. The computer code is
modified to include the evaporation
in the inner disk and the growth of the disk due to redistribution of
angular momentum (Sect. 4-6). We show the results for the disk
evolution  and discuss the influence of parameters in Sect. 7.
The accretion of matter on to the white dwarf releases X-rays. In
Sect. 8 we evaluate how much one may expect to observe and compare
with ROSAT and
earlier observations. In Sect. 9 we discuss modelling with higher
values of $\alpha_c$ in previous work. We point out the consequences from
the fact that the 3:1 resonance radius may be reached several years
before the onset of the outburst. We further present a new view of
Smak's (1993) interpretation of white dwarf luminosity variation,
arising from our results. In the conclusion we sumarize what follows
from our computations: (1) on the low viscosity, (2) on  the importance of
evaporation and (3) on the late evolutionary state of cataclysmic variables. 

\section{Observational data and derived model parameters}
WZ Sg is the dwarf nova with the longest observed outburst period.
The three outbursts recorded in the past, November 22 1913,
June 29 1946 and December 1 1978 occurred after remarkably similar
time intervals of 32.6 and 32.4 years, respectively (Mattei 1980).
WZ Sge is an SU UMa type dwarf nova,
but unique in the way that no normal outbursts occur in
between the superoutbursts. Smak (1993) argued,
that the extremely long outburst cycle can only be explained by 
a very low viscosity in the disk during quiescence. 
From this work we take the following parameters:
 mass of white dwarf $M_1$=0.45M$_\odot$ and  mass of secondary star
$M_2$=0.058 M$_\odot$,  mass overflow rate
from the secondary star of about $\dot M$= 2\,$10^{15}$ g/s. The
amount of matter
in the disk $ M_d$ accumulated until the outburst
should be  1-2\,$ 10^{24}$ g.
We  also consider other parameters to see the
dependence of the results on the assumed values. The white dwarf mass 
is the most important one for the evolution. Sion et al. (1995) and
Cheng et al. (1997) analysed HST observations of WZ Sge and 
from spectra derived values for rotational velocity
and gravity. 
Together with the observed 28\,s period they inferred a white dwarf
mass of 0.3\,M$_\odot$ ( or higher, if the star is differentially rotating).
We add an independent
estimate of about 0.7\,M$_\odot$ from the observed period and assuming
criticical rotation. Spruit and Rutten
(1998) found in their recent investigation 1.2$\pm$0.25$M_\odot$ from an
analysis of the stream impact region.

\section{Previous modelling of the long-time behaviour of WZ Sge}
It is difficult to understand the long outburst recurrence time.
In previous investigations this behaviour was modelled with
either a very small value of the viscosity parameter in quiescence
$\alpha_c$=0.001 or a moderate value $\alpha_c$=0.01, which is only 
slightly smaller than the values usually chosen to describe ordinary
dwarf nova outbursts. 

\subsection{ Model based on $\alpha_c$=0.001}
Osaki (1995) simulated the outburst lightcurve with the the parameters
$\alpha_c$=0.001, $\alpha_h$=0.03 ($\alpha_h$ viscosity in the hot state),
$\dot M$=1.2\, $10^{15}$g/s, $M_1=1M_\odot$ and $M_2$=0.1M$_\odot$
It was assumed that mass is
accumulated in a torus at the circularisation radius (``Lubow-Shu''
radius). This computation was based on the simplifying model of 
Osaki (1989). The observed outburst amplitude and duration are then 
well described. In the thermal-tidal instability model for SU UMa stars 
a decrease of the mass transfer rate leads to a longer
superoutburst recurrence
time and simultanously fewer normal outbursts in between (for a review
see Osaki 1996). According to the standard scenario of CV evolution
(Kolb 1993) mass transfer rates are expected to decrease with age. For the
modelling of WZ Sge at the very end of this
``activity sequence'' not only $\dot M$ has to be low, also $\alpha_c$
must be extremely low.

\subsection{Models based on $\alpha_c$=0.01}
To avoid the assumption of extremely small values $\alpha_c$ two
different attempts were made to understand the behaviour of WZ Sge, both
using a moderate value in combination with a hole assumed in the inner
disk.

Lasota et al. (1995) and Hameury et al. (1997) pointed out that a hole
in the inner disk due to evaporation
into a coronal layer above the cool disk (Meyer \& Meyer-Hofmeister
1994) or due to the presence of a magnetosphere of the white dwarf (Livio \& Pringle
1992) prevents the onset of an outburst in the inner disk. They observe
that the X-ray flux from WZ Sge 
soon after the outburst was roughly the same as several years later
and argue that this points to a
quasi-stationary disk structure during quiescence. Their analysis
shows that with an inner hole reaching to 2.5 - 5\, $10^9$cm,
$\dot M$ =$10^{15}$g/s and
$ \alpha_c$=0.01 no outburst occurs, the disk is stable. 
The
mass contained in the disk is then much less than the value $M_d$
derived by Smak. According to the estimate in Hameury et al. (1997)
it is below
3\, $10^{23}$g/s. To solve the two problems how to get the instability
triggered in the stable disk 
and how to supply a high amount of mass the authors suggest
that a fluctuation in the mass overflow rate from the
companion star might lead to an increase of surface density in the
disk beyond the critical value and trigger the instability. Then a
very high mass overflow is expected due to irradiation of the
secondary star. In their scenario this could
happen any time after the disk is stationary. 

Warner and al. (1996) try to model WZ Sge with the same value
$\alpha_c$= 0.01, a hole of 3\, $10^9$cm , $\dot M$
 = 6\,${ 10^{14}}$g/s, $M_1=0.6M_\odot$ and an outer disk
radius of 1.1\, $10^{10}$ cm. The assumed hole makes it possible to
accumulate mass over a long time until after 36 years the outburst
occurs. But due to the low accretion rate only 7 $10^{23}$ g/s matter were 
transferred from the compagnion to the disk. The computed
outburst lasts only 6 days.

\section{Models with disk evaporation}
Close to
the white dwarf a hot corona above the cool disk leads to evaporation
(Meyer \& Meyer-Hofmeister 1994, F.K. Liu et al. 1995) of the inner
disk. B.F. Liu et al.(1997) have shown in detail  how the
evaporation affects the disk for a typical dwarf nova.
The evaporation rate scales as 
${M_1}^{2.34} $ and $r^{-3.17}$ ( F.K. Liu. et al. 1995). This allows
to evaluate the coronal flow rate and the mass exchange between cool
disk and hot corona (evaporation and condensation) as a function of radius.
In a system like VW Hydri a hole is formed after the end
of an outburst extending to about 2
$10^9$cm and it remains that size during the quiescence until the next
outburst starts.

 The position of the inner edge of the disk is determined by the
balance between evaporation of gas from the cool disk into the corona
and the supply of new matter from outer disk regions. The 
mass flow rate in the disk $\dot M(r)$ decreases with decreasing r.
($\dot M(r)$ is related to the surface density; for a description of disk
structure see Ludwig et al. 1994). Because of this there is always a
position where the rates are equal.
The hole is larger the smaller $\dot M(r)$ is.
The cool disk in WZ Sge relaxes to a quasi-stationary inward flow of
mass due to the long time available in quiescence and the hole can be
closed after a few
years depending on the white dwarf mass. But evaporation is going on
all the time and, at the inner edge, puts a significant fraction of
this flow into the form of hot coronal gas.  

\section{The redistribution of angular momentum in the disk}
The outer edge is determined by the redistribution of angular momentum.
The gas in the incoming stream from the secondary star has a certain specific
angular momentum. The radius at which the gas
in the Keplerian orbit around the white dwarf has the same specific
angular momentum is called ``Lubow-Shu'' radius.

Generally, in dwarf nova systems, there is no significant accretion on
the white dwarf, all mass is kept in the disk during the
comparatively short duration of quiescence. With the addition of low
specific angular momentum gas from the secondary the outer disk radius
shrinks towards the ``Lubow-Shu'' radius, only to expand again during 
outburst due to the flow of matter towards the white dwarf. The
situation in WZ Sge and comparable systems is different. 
During the long quiescence quasi-stationary
accretion on to the white dwarf is established. The specific angular
momentum of this accreted matter is stored in the disk and as a
consequence the disk grows.
This goes on until it reaches the radius where the 3:1 resonance
between Kepler binary period and the
Kepler rotation occurs (which for these low mass ratio $M_1$/$M_2$
systems is inside the tidal truncation radius). Thus the tidal
instability (Whitehurst 1988, Lubow 1991) sets in and the excentric disk produces superhumps in the
lightcurve as discussed by Lubow (1994).
To follow the evolution of systems with a long lasting quiescence like 
WZ Sge the radius change has to be taken into account in the computations.
Ichikawa \& Osaki (1992) had already included in their numerical code 
radius changes for the evolution of U Gem. 

\section{The numerical code}
In our computations we solve the diffusion equation for mass and
angular momentum flow and take the
dependence of the critical values of surface density and the viscosity-
surface density relation as described in Ludwig et al.
(1994). In addition we take into account that the
positions of both inner and outer edge
of the disk change during the evolution.

We use the following formulae. $\dot M_0$ (taken positiv inward),
$\Sigma$ and $\Omega$ are mass flow rate, surface density and Kepler
frequency in the disk, $r_{out}$ is the outer disk radius and index
``out'' designates values at $r_{out}$, index ``LS'' at the Lubow-Shu
radius $r_{LS}$.
Conservation of mass and angular momentum at ${r_{out}}$ give 

\begin{equation}
\dot {M}_0 + \dot {M}_{out} = 2\pi \left( r \Sigma \right)_{out}
 \frac {{\rm d}r_{out}}{{\rm d} t }
\end{equation}

%\begin{equation}
\begin{eqnarray}
{\dot M}_0({{r^2}{\Omega}})_{LS}= \left( (\dot {M}-3\pi
f) r^2 \Omega \right)_{out}+   
\nonumber \\   \left(2\pi r \Sigma r^2
\Omega \right)_{out} \frac {{\rm d}r_{out}}{{\rm d} t },
%\end{equation}
\end{eqnarray}
where  ${\dot M}_{out}$ is the mass flow in the disk at ${r_{out}}$
(taken positive inward). 
Further $f=\int\limits_{-\infty}^\infty \mu dz$ viscosity integral,
$z$ the height
above the midplane, $\mu$ the effective viscosity, parametrized proportional to the
value $\alpha$ (Shakura \& Sunyaev 1973, compare Ludwig et al. 1994). 

In the computer code we have an equidistant grid in $x=2\sqrt{r}$ and
use $b=\sqrt{r}f$ instead of $f$.
 Conservation
of mass and angular momentum in the disk are guaranted with the proper
determination of the position of the outer edge $r_{out}$ and the appropriate
value $b_{out}$ there. $b_{out}$ and $\frac{{\rm d}x_{out}}{{\rm d}t}$
are then determined by Eqs. (1) and (2).
\begin{equation}
b_{out}=x_{out} \frac{\dot{M}_0}{6\pi}(1-x_{LS}/x_{out})
\end{equation}

\begin{equation}
\frac {{\rm d} x_{out}}{\rm dt} = \frac{4}{\pi x_{out}^3 \Sigma_{out}}
\left[\dot{M}_0-6\pi(\frac {{\rm d}b} {{\rm d}x})_{out}\right] 
\end{equation}

$x_{LS}$ is related to $r_{LS}$ (we took the
values of $r_{LS}$
from Table 2 in the investigation of Lubow \& Shu (1975)).

If the 3:1 resonance radius is reached a further growth of the disk is
effectively cut off since the tidal instability quickly increases to such
a strength, that all surplus angular momentum is transferred back to the
binary orbit. We implement this by replacing Eqs. (3) and (4)
by
\begin{equation}
\frac{{\rm d}b_{out}}{{\rm d} x}=\frac{\dot{M}_0}{6\pi}
\end{equation}
and
\begin{equation}
\frac{{\rm d}r_{out}}{{\rm d}x}=0
\end{equation}

\section{New computation of the disk evolution during quiescence}
We follow the evolution of the cool disk from the early quiescence to
the onset of the next outburst.
The orbital period is known. The geometry of the binary system
containing white dwarf and Roche lobe filling secondary star is then
determined, if we assume white dwarf mass. Further parameters
are the mass overflow rate from the secondary star and $\alpha_c$.
We start our investigation with the parameters for $M_1$ and $M_2$ suggested by
Smak (1993). We want to model the evolution that means find the onset of an
outburst after 32 years. We take as constraints the amount of matter 
in the disk at this time, 1-2\, $10^{24}$g, and the mass accretion rate
of 2\, $10^{15}$g/s , also derived by Smak. Note that this accretion
rate will lead to the accumulation of the right amount of matter
during the 32 years, but whether an outburst occurs at this time
depends on the disk evolution, which is governed by the
viscosity. Evaporation is an 
additional important feature, which influences the location where the
instability is triggered. In the following we show the
results,  (1) the time
until an outburst arises and (2) the amount of matter accumulated
in the disk until then. We take $ \alpha_c$=0.001 fixed and
vary the mass accretion rate. Then we discuss the variation of these
quantities with $ \alpha_c$ and the influence of evaporation. Finally we show
results for a  larger white dwarf mass.

% 1
% von msb:\begin{figure}[ht]
% "includegraphics" bindet Postscript-Files ein:
% \includegraphics[width=\hsize]{fig1.eps}
% eps=encapsulated ps-file, muss nicht sein
% und so kann man freien Platz lassen:
% \vspace{5cm}
% \caption{This picture is to show that I can draw like Tizian. Just
%  details have to be filled in.\label{fig:1}}
% \end{figure}
%     alt:\centerline{\psfig{figure=fig1.ps,width=8.8cm}
%          \psfig{figure=fig1.ps,width=8.8cm}}
\begin{figure*}[[ht]
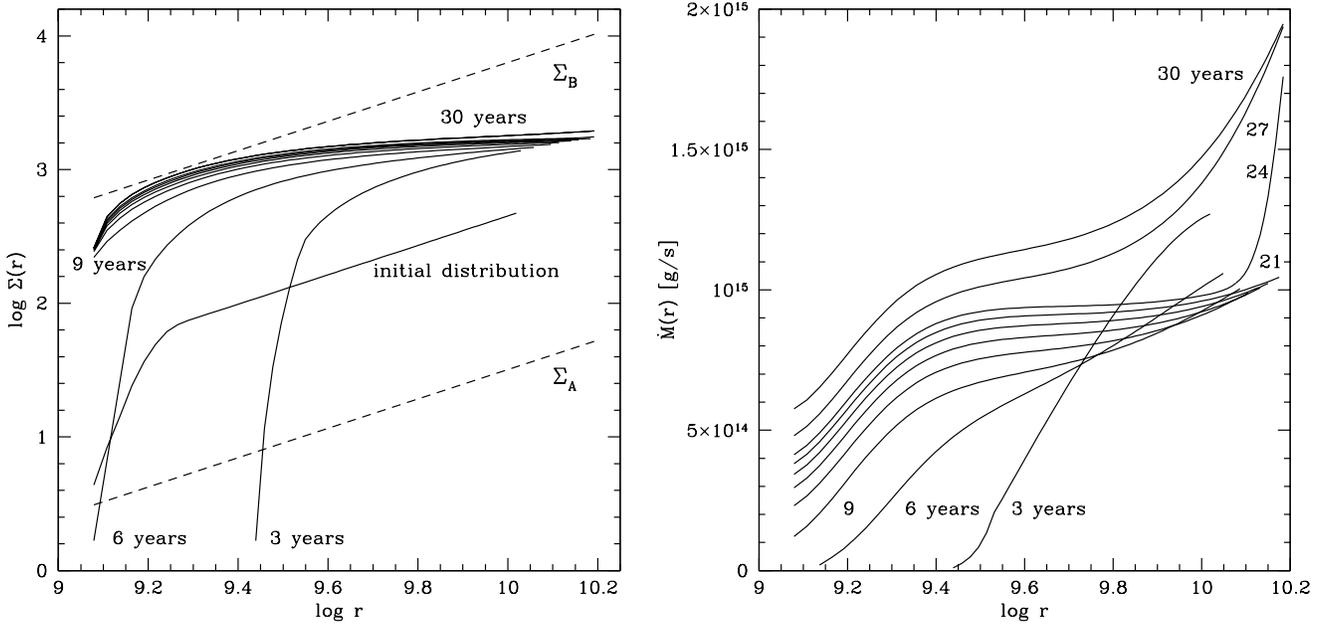

\includegraphics[width=8.8cm]{1051.F1a}
\includegraphics[width=8.8cm]{1051.F1b}
\caption {
Evolution of the disk during quiescence. Left panel: Surface density
$\Sigma$ given at
time intervals of 3 years. The radial growth of the disk can be seen from the
extent of the surface density lines. ${\Sigma_A}$ and ${\Sigma_B}$ are
critical surface densities; above ${\Sigma_B}$ only hot state
possible, which means the outburst starts, if it is reached; r
distance to white dwarf.
Right panel: Corresponding mass flow rates in the disk $\dot M(r)$ }
\end{figure*}

\subsection{Initial distribution}
We take an initial distribution of
surface density in the disk according to the results for ordinary
dwarf  novae of Ludwig \& Meyer
(1998), but scaled to the lower values of
$\alpha_c$ used here.
To start the computation of the long-time evolution we have
to assume the disk size in the beginning. To test the effect of this
we study two cases. In case (a) we
take a relatively small disk. Such a disk then contains not much
mass  ($\approx$ 0.1\, $ 10^{24}$g) and angular
momentum. As an alternative, case (b), we start with a disk reaching
almost to the
resonance radius. Then more mass is in the disk ($\approx$
0.28\, $ 10^{24}$g). Consequently more
matter should be in the disk at the onset of the outburst, so that
the difference in mass $\Delta M_d$ corresponds to the amount
1-2\,$ 10^{24}$g derived from the outburst luminosity.
The total amount accumulated should be higher in case (b) and the
matter has to be packed more densily. This demands a smaller viscosity $
\alpha_c$. All computations if not stated otherwise are case (a).
 The results for
the computations case (b) ($ \dot M $=2\,$10^{15}$g/s,
$\alpha_c$= 0.001 and 0.0008) are marked by the two asterisks in Fig. 3.
Comparing the evolution in case (a) and (b) we find,
that the onset of the outburst happens for the same total amount of matter
in the disk and at nearly the same time. 
An exact initial distribution could only be
determined by a detailed computation of the outburst.
Due to the long
quiescence the evolution is only slightly influenced by this uncertainty.

\subsection{ Results for $M_1=0.45M_\odot$ and $ \alpha_c$=0.001
(``standard parameters'')}
In Fig. 1 we show the evolution
of the disk for our ``standard'' parameters.
After 30.3 years the critical surface density $\Sigma_B$ is reached at
the distance r=1.86\, $10^9$ cm from the white dwarf and the outburst starts.
In the disk 1.44\, $10^{24}$g matter are accumulated.
In the left panel we show the surface density $\Sigma(r)$ in the disk, in
the right panel the corresponding rates  $\dot M(r)$ every 3 years
during the evolution. The run of ${\dot M}$(r) during the first
3 years is due to the assumed initial distribution of surface
density. In the early evolution the
mass flow rate in the inner disk is small and evaporation creates a
hole. The maximal extent of the hole r=3\, $10^9$ cm, is already reached
after less than 3 years.
The position of the inner edge depends on the balance between
the rate of mass flow inward and evaporation. The surface density
increases continously during quiescence and with it $\dot M(r)$.
The hole is closed after 9 years of disk evolution. Evaporation is still
going on, then with the high rate of $\approx$ $10^{15}$g/s. About 20\% of
this flow goes into wind loss, the remaining part accretes via the
hot corona on to the white dwarf.

After the
first decade we find an almost constant mass flow rate of 7-8\,
$10^{14}$ g/s over a large part of the disk. The rate of evaporation,
efficient in a ringlike region around the hole,
can also be seen from the Fig. 1, left panel. It is the difference between the
nearly constant value and the lower values at distances r less than
about 2.5\, $10^9$ cm. The values at the inner disk edge indicate how much
mass flows via the cool disk on to the white dwarf. Differences resulting from 
accretion of hot versus cool gas are discussed in Sect. 6.
The outer radius
of the disk increases as can be seen from the extent of the lines at
different evolutionary times. In Fig. 2 we show the changes of the
outer and inner disk
radius for ``standard'' parameters, but an initially larger disk
(initial distribution case b),where the shrinking of the outer edge is more
prominent. The disk grows until
the resonance radius r=1.6\, $10^{10}$cm ( ``standard'' parameters)
is reached.

The size of the disk is an interesting feature. Before the 3:1 resonance
radius is reached roughly half of the newly accreted matter is
needed to fill the larger and larger outermost disk areas. The
remaining part is too low to trigger an outburst. After the resonance
radius is reached the disk 
does not grow anymore, the mass flow in the disk suddenly is nearly
doubled, and the surface density increases to the critical value.
In Fig. 1, right panel, we see
the higher mass flow rate after the resonance radius is reached. We 
point out that evaporation prevents instability in the
early evolution. Without evaporation the critical
surface density would have been reached at r= 1.5\, $10^9$ after 9.6
years (compare Fig. 3). 
The fact that the hole is closed does not change this situation.

%  2
\begin{figure}[[ht]
\includegraphics[width=7.5cm]{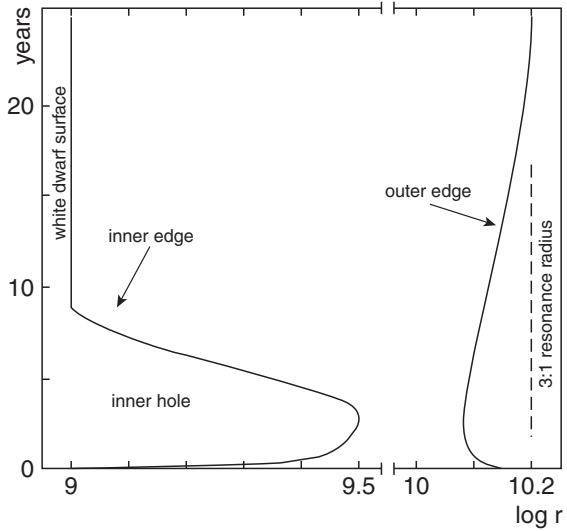}
\caption { Changes of the position of outer and inner disk edge
during the evolution for ``standard case ``
parameters, but a large disk initially.} 
\end{figure}

%  3
\begin{figure*}[[ht]
\includegraphics[width=12cm]{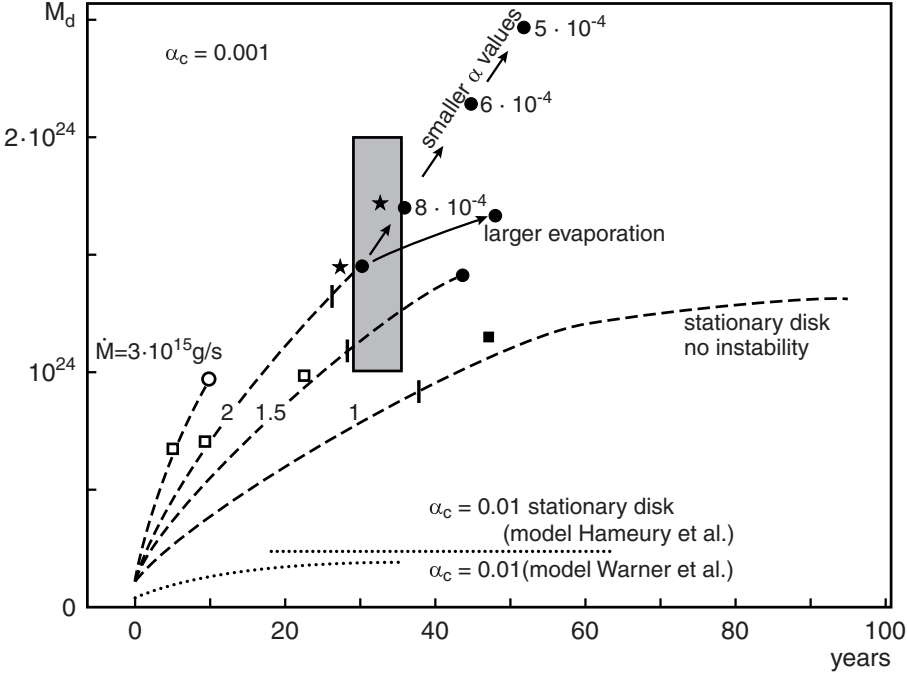}
\caption {
Amount of matter accumulated in the disk after time t for 
mass overflow rates from the companion
star $ \dot M$=1,\,1.5,\,2 and 3\,$10^{15}$g/s. Broken lines describe
the accumulation of matter during the evolution. The shaded box
marks the observed recurrence time and the amount of matter
derived by Smak (1993), which we want to model. 
Circles and squares indicate the onset of an
outburst, either including evaporation or not; bars at the
evolutionary lines mark the
time, when the 3:1 resonance radius is reached; circles/squares are
filled if this is the case, otherwise open. The 2 asterisks show the
results based on initial distribution case (b), to be compared with
the filled neighbouring circles (see
Sect. 7.1); arrows indicate the effect of variation of $\alpha_c$ and
evaporation.
$M_1$=0.45$M_\odot$,
$M_2$=0.058$M_\odot$ . 
Models by Hameury et al. and Warner
et al. (see text) indicated by dotted lines.}
\end{figure*}

If we follow the disk evolution for given parameters $M_1$, $M_2$
and $\alpha_c$ the accumulation of mass depends further on $ \dot M$.
To get a model appropriate for WZ Sge the critical surface density
has to be reached after $\approx$ 32 years and the amount of matter
accumulated then has to be 1-2\, $10^{24}$ g.
Fig. 3 shows this 
accumulation process for different rates of $ \dot M $. Higher rates
make the outburst happen too
early. The opposite happens for low rates. With $ \dot M $=$10^{15}$
g/s, which is half of what was
suggested for WZ Sge (Smak 1993) the surface density in the disk never
reaches the critical value, no outburst occurs. The disk is
stationary. All matter accreted from the secondary star flows through
the disk on to the white dwarf (except the part lost by the wind). This means,
that mass is accreted continously via the corona and the disk on to
the white dwarf. In Fig. 4 we show how the disk reaches such
a stationary state.

For comparison we have also performed computations neglecting
evaporation. As can be seen from Fig. 3 the outburst would happen much
earlier, after 9.6 years instead of 32 years for the same parameters.
 The amount of accumulated matter would be lower than what
corresponds to the observed increase of 7 magnitudes in the outburst
lightcurve and the duration longer than 1 month (Mattei 1980).
Such an outburst is prevented by the formation of a hole.

%   4
\begin{figure}[[ht]
\includegraphics[width=8.8cm]{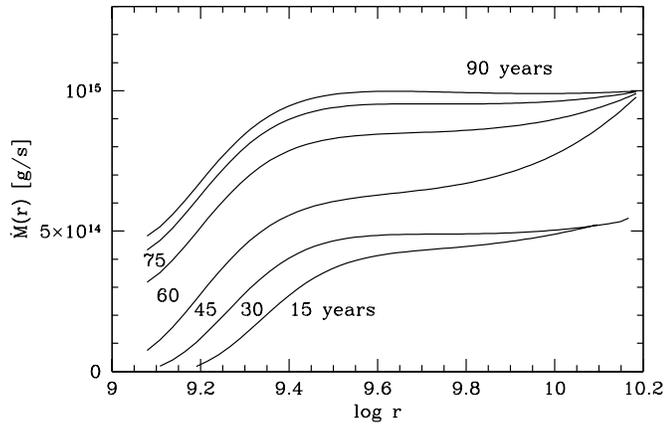}
\caption { Mass flow rates $ \dot M(r)$ in the disk for 
a mass overflow rate from secondary $ \dot M$=$10^{15}$g/s,
otherwise ``standard'' parameters.}
\end{figure}

\subsection{ Results for smaller $ \alpha_c$}
A smaller viscosity means that the mass flow in the disk is reduced
and the matter is packed more densily. For the evolution this results
in a higher $M_d$ and a longer recurrence time. Figures 3 and 5
show the results
for viscosity values $\alpha_c$ = 5-8\, $10^{-4}$ instead of 0.001.

%   5
\begin{figure}[[ht]
\includegraphics[width=7cm]{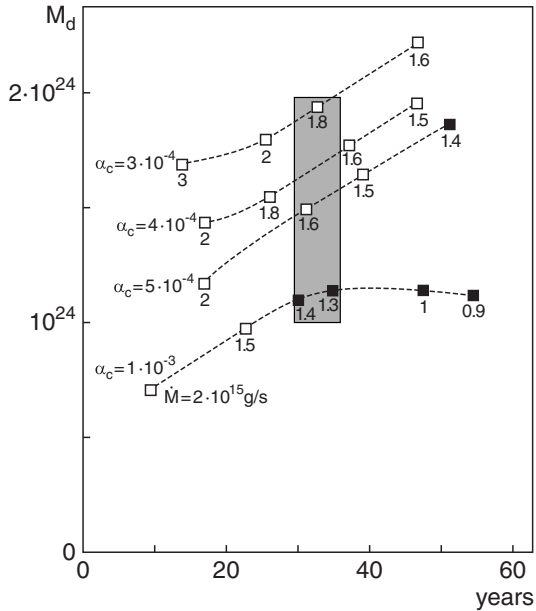}
\caption {Observed recurrence time and amount of matter in the disk
 at the onset of the outburst (shaded box). Squares are the computed
accumulated mass and recurrence time at the onset of the outburst, without
evaporation, as functions of cool state viscosity $\alpha_c$ and 
mass transfer rate $ \dot M$.}
\end{figure}

\subsection{ Results for different strength of evaporation}
We have computed one example of evolution for our ``standard''
parameters, but with an evaporation rate three times as high
as the
regular rate. This means a larger hole, a later onset of the outburst
and more matter accumulated (see Fig. 3).

The opposite case
is neglecting evaporation. We indicated already in Fig. 3 (squares) at which
time the instability would be triggered without a hole in the inner
disk. To fullfill the constraints, long recurrence time and large
amount of matter in the disk, other parameters have to be taken.

We have performed a series of
computations of disk evolution varying $\alpha_c$ and  $ \dot M$. 
The critical surface density is reached the earlier the higher the
mass accretion rate. One needs small
values for $\alpha_c$=
 3\, $10^{-4}$ to
 1\, $10^{-3}$ in combination with 
low rates $ \dot M$= 1.8 to 1.3\, $10^{15}$ g/s to get
the right recurrence time and amount of matter in the disk. 
Figure 5 shows the results in terms of the mass
accumulated at the onset of outburst. Only for the lowest rates $ \dot M$
is the resonance radius reached before the outburst starts. In the
other cases the disk expands to the 3:1 resonance radius during the
outburst. Then superhumps 
would only appear in the outburst. Figure 6 shows the mass flow rates
in the disk for various times during the evolution for a low mass
transfer rate, 1.3\,$10^{15}$g/s,
from the seccondary
star. The disk is quasi-stationary. Without
evaporation the matter is accreted via the cool disk on to the white
dwarf. For a rate a bit lower the critical surface density would never
be reached. So this case is similar to a quasi-stationary disk with
evaporation (see Sect. 7.2, evolution for $ \dot M$=1
$10^{15}$g/s in Fig. 3). 

%   6
\begin{figure}[[ht]
\includegraphics[width=8.8cm]{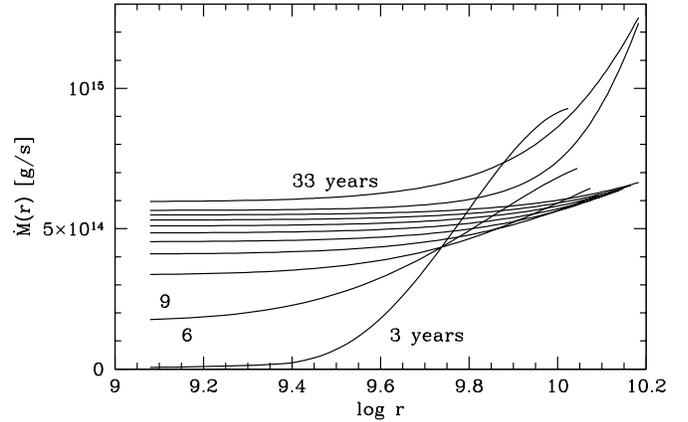}
\caption {Mass flow rates $ \dot M(r)$ in the disk for a mass
accretion rate from the secondary $ \dot M$=1.3\, $10^{15}$g/s
otherwise ``standard'' parameters, but without evaporation}
\end{figure}

\subsection{ Results for larger white dwarf mass}
For larger white dwarf mass the evaporation process is more important.
The higher rate keeps the hole open longer. We computed the evolution for
$M_1=1M_\odot$ ,
$M_1=1.2M_\odot$ and the mass ratio $M_2/M_1$=0.075. For the larger
white dwarf mass a larger fraction of
matter passes continously
through  the disk and in the inner part 
through the corona on to the white dwarf. Less matter can be
accumulated in the disk. We found the following results.
 (1) $M_1=1.2\,M_\odot$,
 $\alpha_c$=0.001, $ \dot M$=2\, $10^{15}$g/s (rate ``standard parameters'')
no outburst. (2) $M_1=1M_\odot$, $\alpha_c$=0.001, $ \dot M$=2\, $10^{15}$g/s, 
outburst after $\approx$100 years, this means this rate corresponds to
an almost stationary disk. If we increase $ \dot M$ to 3 $10^{15}$g/s an
outburst arises after 50 years. In all three examples $\approx$3
$10^{24}$g of matter were accumulated. In agreement with our findings 
from parameter variations described earlier we get for the 1$M_\odot$
white dwarf about
the right recurrence time 29.2 years and amount of matter 1.36 $10^{24}$g 
with $\alpha_c$=0.003 and $ \dot M$=2.5 $10^{15}$g/s. The mass flow rate
on to the white dwarf increases to values around 2 $10^{15}$ during the
quiescence. This and the deeper gravitational potential of the more
massive white dwarf implies a
higher X-ray flux from the system.

\section{X-rays}
\subsection{Observations for WZ Sge}
Observations of X-rays from cataclysmic variables (CVs) with the
Einstein Observatory also included WZ Sge (for a review see Cordova \& Mason
1983, for the determination of temperatures Eracleous et al. 1991).
WZ Sge was observed in April 1979, April and May 1980. It would be
interesting to learn about changes shortly after the outburst.
The low number of counts had been at about the
same level for the three times of observations, but the derived
best fit temperatures variied between 2.9 and 5.8 keV. The X-ray flux
(0.1-3.25 keV), unabsorbed, was 0.46 to 0.62\, $10^{-11}$ ${\rm {erg
 cm^{-2}s^{-1}}}$.
Mukai \& Shiokawa (1994)
analysed the EXOSAT ME archival data on dwarf novae.
From the brightest observation of WZ Sge in October 1985 they found the X-ray
flux (2-10 keV) of 0.39\, $10^{-11}$ ${\rm {erg cm^{-2}s^{-1}}}$.
Richman (1996) analysed the
X-ray flux from CVs observed with the ROSAT PSPC.
From the observations of WZ Sge taken in April 1991 she found the best fit
temperature 3.4 keV and the flux (0.1-2.4 keV) 0.13\, $10^{-11}
{\rm {erg cm^{-2}s^{-1}}}$.

Using a correlation between mass accretion rate and the ratio of X-ray to
visual flux (Patterson \& Raymond 1985) the ROSAT observations indicate
a mass accretion rate of {$10^{14.6}$g/s} (Richman 1996). This
distant-independent estimate from the observations supports our
theoretical picture, that accretion goes on all the time. 
The amount of mass flow in quiescence of WZ Sge as suggested by
our model agrees with this estimate.

\subsection{Expected X-ray flux}
Accretion on to a white dwarf at a rate of $10^{-14.6}$g/s releases energy,
 which, if 
mainly in X-rays, would predict significantly higher fluxes at earth
than observed. There are however a number of reduction factors, which
together bring prediction and observation into approximate agreement.

\subsubsection{Accretion on to a slowly rotating white dwarf}

We first discuss the release of accretion energy in the surroundings of the
white dwarf. During the
quasi-stationary evolution the mass flow rate is around
$10^{-11}$$M_\odot$/y in the middle disk (at distances larger than
about $10^{9.5}$cm
, compare Fig. 1, right panel). Farther in,
roughly $\frac {1}{3}$
of the mass flow remains in the optically thick disk, $\frac {2}{3}$
evaporate into the corona and of these, $\approx$ 20\% are lost in a
wind.

 If the white dwarf is not close to critically rotating,
both flows, optically thick and thin, dissipate their rotational
energy, about half of the accretion energy, in a frictional boundary
layer close to the surface of the white dwarf (Pringle 1977, Pringle \&
Savonije 1979). The boundary layer related to the coronal flow is always
optically thin. At these low accretion rates also the boundary layer
related to the flow via the disk is probably optically thin (Narayan \& Popham 1993,
Popham \& Narayan 1995). The other half of the
accretion energy of the coronal flow is dissipated in the corona
above the optically thick disk. What is not used up by the escaping
wind is conducted down and radiated away in the thermal
boundary layer between corona and disk. The energy dissipated in
both, white dwarf and coronal boundary layer adds up to an efficiency 
of about 0.5 of the total gravitational energy release.
With a hole the situation is roughly the same.

Another reduction is expected from the spectral
distribution of X-rays. F.K Liu et
al. (1995) investigated how much X-rays and UV radiation are caused by
the accretion of matter evaporating from the disk and flowing via the hot
corona on to the white dwarf. A theoretical spectrum
was compared with ROSAT observations for VW Hydri
(Meyer et al. 1996). The form of this spectrum is very general due to the
scaling properties of such transition layers. Fig. 7 shows the
luminosity  contributions of the various temperature layers in the
thermal boundary layer of an 
accreting white dwarf. 
The multi-temperature character clearly cannot be satisfacturely be
described by a one-temperature spectrum (see
also Richman 1996). Fig. 7 shows that a fraction of about 0.4 of
the radiation is below the sensitivity limit of 0.1 keV of the various X-ray
observatories. For ROSAT also an upper cut-off might matter.

%  7
\begin{figure}[[ht]
\includegraphics[width=7.5cm]{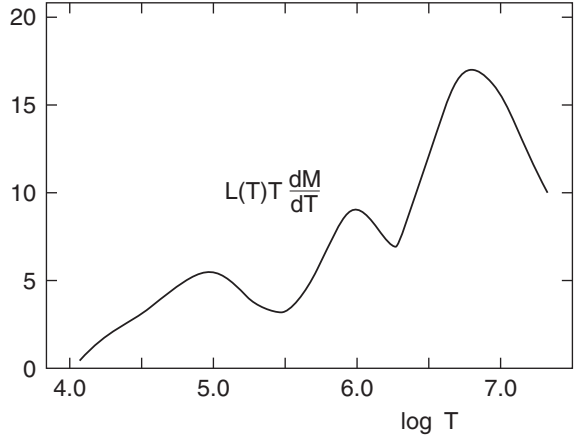}
\caption {Luminosity contributions of the
various temperature layers in the thermal boundary of an 
accreting white dwarf,
T temperature in Kelvin, M emission measure and L(T) luminosity per
andunit emission measure of the
optically thin plasma in non-dimensional
units (according to F.K Liu et al. 1995). The area underneath the
curve  gives the contribution of the
corresponding temperature interval. Note the multi-temperature
character and the significant contributions below 0.1 keV ($\approx$
 1.2\, $10^6$K).}
\end{figure}

A further reduction of the observable flux is due to the fact that
half of the optically thin emission layers are hidden by the
white dwarf and accretion disk surfaces, respectively. 
The reduction is 0.5 for a disk with an extended hole. If the disk
reaches near to the
surface of the white dwarf a further 
reduction results, because the white dwarf
hemisphere towards the observer is partially
covered by the disk and vice versa, depending on inclination. 
The total observable fractions are then 0.32 for the white dwarf and
0.4 for the disk (i=75$^\circ$ for WZ Sge, Smak 1993).

These reductions add up to 0.15 and 0.10 with and without an extended
hole. This gives an observable X-ray luminosity of
\begin{eqnarray}
 L_{\rm{x,obs}}= 0.10 \frac{GM_1 \dot{M}}{R}
\end{eqnarray}
\noindent
R radius of white dwarf. For $M_1=0.5M_\odot$ the flux density at earth
(unabsorbed) is 
\begin{eqnarray}
 f{\rm}_x & = & {L_{\rm {x,obs}}}/ {4\pi d^2} \nonumber \\ 
     & = & 9.7\times 10^{-12} \dot {M}_{-11}
\left( \frac{d}{60\rm {pc}}\right)^{-2}
\frac{\rm erg}{\rm{ cm^{2} s}}
\end{eqnarray}
\noindent
$\dot{M}_{-11}$ is the mass accretion rate in units of $10^{-11}
 M_\odot$/y.
This is about double the observed flux.

\subsubsection{WZ Sagittae, a critically rotating white dwarf}

Now we discuss the consequences of critical rotation on the observable
X-ray flux and the white dwarf mass.

The photometric period of 28s discovered in UV light from WZ Sge by
Welsh et. al. (1997 and references to earlier work there) appears to
originate at the surface of the white dwarf and indicates a very rapid
rotation of this star (see also Cheng et al., 1997). Using the
mass-radius  relation and the variation of equatorial radius with rotation
interpolated from Hachisu (1986) by Popham \& Narayan (1995) we can
determine which white dwarf rotates critically at its equator with
this period. We find
\begin{eqnarray}
 M=0.69 M_\odot,
 \,R_{\rm{equ}}= 1.2 \times  10^9 {\rm {cm}}
\end{eqnarray}
\noindent 
where $R_{equ}$ is the equatorial radius of the critically rotating
star, 1.6 times that of a non-rotating white dwarf with the same
mass. This mass is a lower limit since with this period any less
massive white dwarf with its larger radius and smaller gravity would
rotate supercritically.
Though the mass of the white dwarf in WZ Sge might be larger and
the star then rotate subcritically it is fairly probable that WZ
Sge has already reached critical rotation during its  long
evolution. One can evaluate how much mass transfer to the white dwarf
is required to speed it up to critical rotation. For example, if one
assumes that mass is accreted with Kepler angular momentum at the
equator and that all accreted matter is lost again in consecutive nova
explosions but with the mean angular momentum
of the white dwarf surface then one obtains critical rotation after
mass transfer of
\begin{eqnarray}
\Delta M=\frac{3}{2}\,{r_g}^2\, \ln \,3 \, M_1= 0.21  M_\odot
\end{eqnarray}
\noindent
where the radius of gyration $r_g$ for a $0.69 M_\odot$ white dwarf,
$r_g^2=0.184$ (H. Ritter, private communication) was used. If WZ Sge
entered the cataclysmic variable evolution with an orbital
period above 3 hours this amont of mass has already been transferred and
the white dwarf will now be critically rotating.

The X-rays from accretion on to a critically rotating white dwarf are
even more reduced since no frictional boundary layer
forms. Material from an optically thick accretion disk then settles on
the  white dwarf surface without becoming optically
thin and producing X-rays. The only remaining contribution comes from the
coronal gas that releases 1/2 of its
gravitational energy as it spirals towards the white dwarf
surface. Again roughly $20\%$ will be lost in a wind together with
about twice its escape energy. The remaining
energy release is then $\frac{1}{8} GM_1 {\dot M}_c/R$.

The coronal accretion rate $\dot{M}_c$ is 0.8 of the mass flow rate
that has disappeared from the flow of the optically thick disk on
approaching the white dwarf, in our example of Fig. 1, right panel,
typically $\dot{M}_c\approx 0.5 \dot{M}$.
Of the energy released in X-rays 1/2 is covered by the disk surface
(or by the white dwarf if part is conducted to the white dwarf
surface), about 0.2 is further hidden behind the white dwarf for WZ Sge's
inclination, and 0.4 is out of spectral range (compare Sect. 8.2.1).
Alltogether this gives an observable X-ray luminosity of
\begin{eqnarray}
 L_{{\rm x,obs}}=0.03  \frac{GM_1 \dot {M}_c}{R}\approx
 0.015 \frac{GM_1 \dot{M}}{R}
\end{eqnarray}
and a flux density at earth (unabsorbed) of 
\begin{eqnarray}
 f_{\rm x} & = & {L_{\rm{x,obs}}}/ {4\pi d^2} \nonumber \\ 
     & = & 1.6 \times 10^{-12} \dot {M}_{-11}
\left( \frac{d}{60\rm {pc}}\right)^{-2}
\frac{\rm erg}{\rm{ cm^{2} s}}
\end{eqnarray}
\noindent
for $M_1=0.69M_\odot$ and R the equatorial radius.
\\
These estimates show that the accretion rates in quiescence determined by our models
lead to X-ray fluxes in good agreement with what has been observed,
1.3 to 6.2   $10^{-12} {\rm erg  cm^{-2}s^{-1}}$, resolving an apparent
puzzling discrepancy. A slight preference for a critically rotating
white dwarf is indicated.

\subsubsection{Magnetic accretion spots on the white dwarf?}
The pulsations observed in various wavelength ranges might suggest
accretion on the rotating poles of a magnetic white dwarf. Such a field would
have to be weaker than $10^4$ gauss. Otherwise the disk/magnetosphere
interface reaches beyond the co-rotation radius and matter entering
the magnetosphere would be flung out instead of accreted.

However, the only intermediate quality of the optical pulsations,
$Q\simeq 10^5$, the fact that the mean period varied by 1.3\,\promille
~between 1995 and 1996 (Patterson et al. 1998), and the difference
between the UV-period (28.10s, Welsh et al. 1997) and the X-ray/
optical period (27.86s, Patterson et al. 1998) argues against an
accretion spot frozen into the rotating white dwarf. It might rather
suggest accretion spots formed by magnetic fields brought in with the
accreted matter and circulating around the nearly critically rotating
white dwarf, similar to what has been suggested for dwarf nova
oscillations (Meyer 1997). Optically thin accretion at higher
latitudes (X-rays, also reprocessed into optical light) could have a
slightly shorter period than equatorial, optically thick disk
accretion (UV light). 

The situation remains unclear and needs further study. The magnetic
pressure of a dipolar white dwarf field of surface strength $10^4$
gauss could just interfere with coronal evaporation pressures at
$r= 10^{9.4}{\rm cm}$ and affect the coronal flow, possibly reducing
the X-rays even below the observed level. Accretion along the magnetic
funnel of accretion formed accretion spots would concentrate the
thermal boundary layer to a smaller surface area and result in an
increase of the maximum temperature $\sim ({\rm area})^{-2/7}$. The above
estimates of X-rays observable will not be much affected by this.

\section {Discussion}
\subsection{Comparison with previous modelling}
We shortly discuss here the earlier modelling described in Sect. 3 in
the context of our results.
\subsubsection{Model of Osaki (1995)}
The simplifying model made already clear that low viscosity values
are needed. Our more detailed computations allow to study two effects,
which are important for the evolution, (1) the amount of matter not
accumulated in the disk, because it is accreted on to the white dwarf
or lost in
the wind from the corona during quiescence; (2) the growth of the disk
with respect to the 3:1 resonance radius.

\subsubsection{Model of Lasota et al. (1995) and Hameury et
al. (1997)}
In this model the disk, with an assumed inner hole of specified radius,
is stationary and stable. 
A fluctuation in the mass overflow rate is
then thought to bring the disk to the point of instability and trigger
the outburst. One would then expect quite a fluctuation in repetition
times
between consecutive outbursts. This contrasts with the nearly exactly
(only 2\% variation) regular repetition observed and the very regular build-up
to outburst in the model presented in this paper. A further problem
might lie in the (presumed irradiation caused) large mass transfer
required during outburst. It becomes necessary since the assumed
stationary disk contains too little mass to explain the observed
outburst energy. It would require an increase in the overflow rate by
a factor of about 300 while observed increases in some dwarf novae
are limited to only a factor 2 (Smak 1995, Hameury et al. 1997).

\subsubsection{Model of Warner et al. (1996)}
Also in this model a hole is assumed. We note that here as in the
preceding model, a more detailed physical picture of the hole
formation and change of its size can have a significant effect on the
duration of the quiescence.
The chosen low mass overflow rate from the secondary makes it
possible to have the outburst occur after only 36 years. But
the low amount of matter provided for the outburst then shows up in the
computed short duration. The authors suggest
that in SU UMa systems small, regular outbursts are followed by larger
superoutbursts. But for WZ Sge such regular outbursts were not observed.
It seems that the claim ``down with low $\alpha$'' 
cannot be realized as suggested.

\subsection{Re-interpretation of the observed UV flux of WZ Sge after the
outburst in the context of our results}
Smak (1993) used IUE observations of WZ Sge, taken 1979 July 11 and 1981
November 22, 0.6 and 2.4 years after the December 1 1978 outburst
to determine temperatures. Together with observed fluxes he derived
values for the product $x(R/d)^2$, where x is the fraction of the full
``stellar disk'' $\pi \,R^2$ that is not
occulted by the accretion disk and is radiating with the effective
temperature determined. His values are 2.72\,$10^{-23}$ and
1.56\,$10^{-23}$, respectively.
Smak assumed ${\rm x_{79}}$=0.63 (${\rm x_{79}}$ 
value in 1979, ${\rm x_{81}}$ in 1981)
corresponding to a full disk whose lower half is partially
occulted by the accretion disk (i=75$^\circ$). From the ratio of the
derived values $x(R/d)^2$  one gets ${\rm
x_{81}}$=0.35. This was interpreted as shrinking of the white dwarf
radiating area to an equatorial belt. The long-term IUE observations
indicate a cooling of the white dwarf (la Dous 1994).

Our model allows another interpretation. If in the 1979
observation an inner disk hole was present as expected theoretically 
in the early
phase of quiescence, flux from one full hemisphere of the white is
observable and, in Smak's terminology, the fraction ${\rm x_{79}}$ is
equal to 1. From the ratio of the values determined by Smak then follows 
${\rm x_{81}}$=0.59. This agrees, within the errors with 0.63
corresponding to the fraction visible if the hole is closed. This
changes the estimate for the distance to WZ Sge from Smak's value and
error
d\,=\,$48\pm10$\,\,pc to d\,=\,$60.5\pm13$\,\,pc for the $0.45M_\odot$
white dwarf. For
the critically rotating 0.69$M_\odot$ white dwarf one may estimate
(see Hachisu 1986) a radius $\approx$ 
0.77 $R_{\rm{equ}}$ of a
``stellar disk'' of the same area as the
rotationally flattened cross section of the real star. With the
equatorial radius of about 1.2\,$10^9$cm this yields the distance
estimate d\,=\,$50.6\pm11$\,\,pc. A not critically rotating star of larger mass
would be correspondingly closer.

The observed change in the UV flux between 1979 and 1981 supports the
prediction of formation and disappearance of an inner hole of the
quiecent accretion disk.

\subsection{Consequences from reaching the 3:1 resonance radius}
From our computations we found that the 3:1 radius can be reached
already several years before the outburst (in Fig. 3 the
bars at the evolutionary lines mark the time when the resonance radius
is reached). This means that the
outer disk from this time on has an excentric shape and superhumps
should be seen. They dissipate the work done in transferring accreted
angular momentum back to the orbit, $\approx\,(\Omega_{out} -
\Omega_{orb})(r^2\Omega)_{LS} \dot M$, which is about twice the hot
spot luminosity ($\Omega_{orb}$ orbital frequency). We note that the weak
early superhumps in outburst do not differentially precess (Whitehurst
1988, DeYoung 1995, Patterson 1995, Kato et al. 1996) and we might
expect the same stationarity in the lightcurve for such quiescent
superhumps.

\section {Conclusions}
Our detailed computations show that values $\alpha_c$ around 0.001 are
necessary to accumulate enough matter in the disk to explain outburst
duration and amplitude. Our investigation
shows new features which are essential for the understanding of WZ
Sge. The evaporation
plays an important role in the way that it prevents a premature
outburst. As long as the disk size grows, more than
half the matter transferred from the secondary star is used up to fill the
newly added outer disk areas. Only after the disk
has reached the 3:1 resonance radius no further growth occurs and all
accreted matter can flow inward. This results in a higher mass flow
rate and triggers the outburst. Evaporation goes
on all the quiescence with the rate corresponding to the inner edge of the
disk close to the white dwarf. This means accretion of matter via the
corona on to the white dwarf at a constant rate of about $10^{15}$g/s.
Without evaporation the outburst behaviour of WZ Sge can also be
modeled, but then requires
small values $\alpha_c$ for the cool state, $\leq$0.001, and low
mass overflow rate from the companion sta,r of order
$\dot M$=1.4\, $10^{15}$g/s.

The finding that the value of $\alpha_c$ has to be smaller than
usually
assumed for dwarf novae raises again the question of the cause. A new 
interesting suggestion of Gammie \& Menou (1998) relating low
viscosity to low magnetic Reynolds numbers does not seem to explain
why the dwarf novae in late evolutionary phase should have low values 
with otherwise very similar quiescent disks.

Our estimate of the observable X-rays from WZ Sge yields agreement with
the observations and slightly favours a critically rotating white
dwarf of 0.7\,$M_\odot$ at a distance of 50 pc.

Most of our models predict superhumps before the outburst. Their
appearance in late quiescence, observable as a change in
the orbital lightcurve, would signal that the disk has reached its
final radius at 3:1 resonance and that after the disk relaxation time
the next outburst will occur.

Models for WZ Sge are relevant for dwarf novae in a late stage of secular
evolution. WZ Sge is already a marginal system in that an
outburst occurs only after a very long
accumulation time. If the accretion rate would be only slightly
smaller, no outbursts would arise. Instead the disk becomes
stationary. 
Since the mass transfer rate from secondary star to white dwarf 
decreases during the secular evolution of dwarf novae (Kolb 1993) we
expect a large number of systems ``beyond WZ Sge''. Such systems are not
conspicuous optically, but the X-rays from all
these binaries add up to quite an amount of radiation.
Consequences arising from this will be
investigated separately (Kolb et al., in preparation)

The features described for WZ Sge have relevance also for other
systems including X-ray transients (Mineshige et al., to be published 
in PASJ).

\begin{acknowledgements}
We thank Ulrich Kolb, Hans Ritter and Henk Spruit for
valuable discussions.
\end{acknowledgements}

\end{document}